\documentclass[epsfig,graphicx,ndvi,ndvs,twocolumn,aps]{revtex4}

\usepackage{epsfig}
\usepackage{graphicx}
\usepackage{dcolumn}
\usepackage{amsmath}
\usepackage{pstcol,times,color,graphicx,graphics,
pstricks,pst-node,pst-coil}

\begin{document}

\title{ Homoclinic Chaos in Axisymmetric Bianchi-IX
cosmological models with an ``ad hoc" quantum potential}

\author{G.C. Corr\^ea}
\email{gcoelho@if.ufrj.br}
\author{T.J. Stuchi}
\email{tstuchi@if.ufrj.br}
\author{S.E. Jor\'as}
\email{joras@if.ufrj.br}
\affiliation{Instituto de F\'{\i}sica, Universidade Fe\-de\-ral do
Rio de Janeiro, Caixa Postal 68528, Rio de Janeiro, RJ 21941-972,
Brazil}

\date{\today}
\begin{abstract}
In this work we study the dynamics of the axisymmetric Bianchi IX cosmological model
with a term of quantum potential added. As it is well known this class of Bianchi IX models are homogeneous and  anisotropic with two scale factors, $A(t)$ and $B(t)$, derived from the solution of Einstein's equation for General Relativity. The model we use in this work has a cosmological constant and the matter content is dust. To this
model we add a quantum-inspired potential that is intended to represent short-range
effects due to the general relativistic behavior of matter in small scales and play the role of a repulsive force near the singularity. We find that this potential restricts the dynamics of the model to  positive values of $A(t)$ and $B(t)$ and alters some qualitative and quantitative characteristics of the dynamics studied previously by several authors. We make a complete analysis of the phase space of the model finding critical points, periodic orbits, stable/unstable manifolds using numerical techniques such as Poincar\'e section, numerical continuation of orbits and numerical globalization of invariant manifolds. We compare the classical  and the 
quantum models. Our main result is the existence of homoclinic crossings of the stable and unstable manifolds in the physically meaningful region of the phase space (where both $A(t)$ and $B(t)$ are positive), indicating chaotic escape to inflation and bouncing near the singularity.
\end{abstract}
\maketitle
\section{Introduction}
Belinskii, Khalatnikov and Lifishitz \cite{BKL} started the  question of 
chaotic behaviour of general Bianchi IX  models  in Relativistic Cosmology. 
The interest in the chaoticity (or not) of Bianchi IX models has been mainly focused on the Mixmaster case (vacuum Bianchi IX models with three scale factors \cite{misner}). The question of the generic behaviour (chaotic or not) of the Mixmaster dynamics  remained unsettled mainly due to the absence of an invariant (or topological) characterization  of chaos in the model (standard chaotic indicators as Liapunov exponents being coordinate  dependent and therefore questionable \cite{matsas,book}). For discussions of the issue of chaotic dynamics on these models we refer to the works of \cite{barrow, berger1, berger2, rugh, burd}.

In the early literature Hawking and Page \cite{page} have shown that Bianchi IX models 
with a scalar field present a fractal set of uncountably infinite oscillatory orbits, 
which  could be considered a signature of chaos in the dynamics. Cornish and
Levin \cite{cornish} proposed to quantify chaos in the Mixmaster universe by
calculating the dimensions of fractal basin boundaries  in initial-conditions
sets for the full dynamics, these boundaries being defined by a code
association with one of the three outcomes on which one of the three axes is
collapsing most quickly, as established {\it numerically}. Jor\'as and Stuchi \cite{stuchi} examined chaos in FRW models with a coupled scalar field by extending the analysis to the complex plane and found a family of non-collapsing structures. In particular  homoclinic chaos in axisymmetric Bianchi IX universes with matter and cosmological constant has been treated in \cite{stuchi2} using coordinate independent topological structure of the dynamics to characterize chaotic behavior in the Hamiltonian system and its physical implications. This paper follows along these lines.
  
The phase space of the classical model is noncompact and the presence of the cosmological  constant determines two crucial facts in phase space: 
first, the existence of a critical point of the saddle-center type; 
second, two critical points at infinity corresponding to the attractor configuration, 
one acting as an ``atractor" to the dynamics and the other as a ``repeller". 
 With respect to the latter point, this system has mathematically the characteristics 
of a chaotic scattering system with two abosolute outcomes consisting of (i) escape to infinity  or (ii) recollapse to the singularity.  The presence of this critical point is responsible for a rich and complex dynamics, engendering in phase space topological structures such as homoclinic orbits to a center manifold. The physical singularities are the main point in the whole discussion, that is, when any one of the scale factors crosses zero, meaning a recollapse of the universe. As showed in \cite{stuchi2} any homoclinic crossings  present in the dynamics of the classical model is not seen by the physical world since the mandatory recurrence is lost because the physical dynamics has to be restricted to $A(t)$ and $B(t)$ greater than zero. Therefore the only separation between recollapse and escape to the attractor at infinity are the unstable and stable manifolds corresponding to the center manifold associated to the Einstein singularity. This establishes the difference between physical and mathematical integrability: in spite of the chaotic dynamics present in the equations the physical meaningful region does not see it (see also \cite{uggla}).
 
In the present  work we study the dynamics of the Bianchi IX cosmological model
as in \cite{stuchi2} to which we add a term of quantum potential inspired by the work
of Alvarenga et al. \cite{lemos} whose  presence represents exactly the short-range
effects due to the quantum behavior of matter in small scales and plays the role
of a repulsive force near the singularity. In this work a similar term has been
introduced  in an ``ad hoc" manner. As it will be seen, this potential restricts the dynamics of the model to the positive values of $A(t)$ and $B(t)$ and alters  some qualitative and quantitative characteristics of the dynamics of the classical model.
We show the common features of a large class of such potentials which depends only on a so-called $r$-equivalent variable: $(AB^2)^{1/3}$. Picking a particular example,
 we make a complete analysis of the phase space of the model finding critical
points, periodic orbits, stable/unstable manifolds using numerical techniques such
as Poincar\'e section, numerical continuation of orbits and numerical globalization of invariant manifolds. We compare the classical and the quantum models and verify that the addition of this quantum term allows the existence of homoclinic crossings of the stable and unstable manifolds in the physical meaninful region of the phase space (both $A(t)$ and $B(t)$ positive) thus allowing chaotic escape to inflation as well as chaotic bouncing near the singularity due to a new center-center equilibrium point. 

\section{The axisymmetric Bianchi IX models with a quantum potential}

In this section we present the  system and examine the main characteristics
of the phase space of the axisymmetric Bianchi IX with a quantum potencial together
with the classical model so that the distinctions between the two models become apparent. We study the main characteristics of the phase space which can be obtained from linear analysis, such as the  equilibrium points and their nature; we present also the invariant plane which is similar to the classical Bianchi IX model.

 Axisymmetric Bianchi-IX cosmological models are characterized by two scale functions
  $A(t)$ and $B(t)$ with the line element \cite{ryan} 

\begin{eqnarray}
\label{eq1}
\nonumber
ds^2 &=& dt^2 - A^2(t)({\omega^{1}})^2 - B^2(t)\times \\
&&\Big[({\omega^{2}})^2+({\omega^{3}})^2 \Big].
\end{eqnarray}
\noindent Here $t$ is the cosmological time and the
${\omega^i}$ are Bianchi IX 1-forms satisfying $d{\omega^i}={\epsilon^{ijk}}{\omega^j}\wedge{\omega^k}$.
 The matter content of the models is assumed to be a pressureless perfect fluid, 
namely dust, with energy density ${\rho}$ and four velocity field $\delta^{\mu}_{0}$ in the comoving coordinate system used, plus a positive cosmological constant $\Lambda$. The dynamics of the scale factors $A(t)$ and $B(t)$ is given by Einstein's equations, which are equivalent to Hamilton's equations generated by the Hamiltonian constraint
\begin{eqnarray} 
\label{eq2}
\nonumber  
H &=&
\frac{p_{A}p_{B}}{4B} - \frac{A p_{A}^2}{8 B^2}+ 2 A - \frac{A^3}{2 B^2}\\
&-&2 \Lambda A B^2-E_{0}=0.
\end{eqnarray} 
where $p_A$ and $p_B$ are the momenta canonically conjugated to $A$ and $B$, respectively, and $E_0$ is a constant proportional to the total matter content of the model, arising from the first integral of the Bianchi identity: $\rho A B^2=E_0$. 

The  quantum term that we introduce in this paper is inspired by the results
Alvarenga et al. \cite{lemos} for the Friedmann-Robertson-Walker with a single scale factor representing the radius of the universes. These authors study the behavior of the wave functions derived from the Wheeler-DeWitt equations. According to the causal Broglie-Bohm interpretation of quantum mechanics these wave functions are associated to particles whose trajectories approach classical trajectories in the classical limit. However in the regions of quantum validity the behavior differs from the classical one. They found that this difference is equivalent to the addition of an ad hoc potencial term  to the classical model, equivalent to a repulsive force. This term agrees with the behavior of the trajectories in the quantum limit. In the work of Alvarenga et al. the quantum potential  term has the following expression:

\begin{eqnarray}
\label{eq9a}
V \propto \frac{1}{a^2}
\end{eqnarray}
\noindent
where $a$ is the radius of the universes. Since our model has two scale factors $A(t)$ and $B(t)$  with volume of space given by $AB^2$ we have chosen an average radius given by $r\equiv(AB^2)^{1/3}$ and  write the new Hamiltonian as:

\begin{eqnarray} 
\label{eq2a} 
\nonumber
H &=&
\frac{p_{A}p_{B}}{4B} - \frac{A p_{A}^2}{8 B^2}+ 2 A - \frac{A^3}{2 B^2}\\
&-&2 \Lambda A B^2+ \sigma V(r)-E_{0}=0
\end{eqnarray} 
where, as before, $p_A$ and $p_B$ are the momenta canonically conjugated to $A$ and $B$, respectively, and $\sigma$ is a constant. We note that Hamiltonian (\ref{eq2a}) with $\sigma=0$  reduces to the classical one (\ref{eq2}). However  it is clear that the new Hamiltonian is not compatible with Einstein's equations and the quantum term is completely ``ad hoc". We now compare the linear behavior and some other features of both the classical and the modified Hamiltonian.

For $\sigma=0$, the classical case, the corresponding Hamilton's equations have
 an equilibrium point whose coordinates are determined by a very easy equation and are:

\begin{equation}
\label{eq4}
E~: ~p_A = p_B = 0, \quad A_0= B_0=\frac{1}{\sqrt{4 \Lambda}}, 
\end{equation}
with associated energy $E_{0}=E_{crit} =\sqrt{\frac{1}{4\Lambda}}$. The eigenvalues are  ${\lambda^c}_{1,2}=\pm \sqrt{\Lambda}$
${\lambda^c}_{3,4}=\pm 2i \sqrt{2 \Lambda}$ which
characterize the equilibrium point $E$ as a saddle-center. For the value of $\Lambda=\frac{1}{4}$ chosen in this work
we have  ${\lambda^c}_{1,2}=\pm \sqrt{2}$, ${\lambda^c}_{3,4}=\pm i \sqrt{2}$ at $A_0=B_0=1$.

When the quantum term is added, the equation for the equilibrium points has to be determined numerically for each pair of parameters $(\Lambda,\sigma)$. Nevertheless, we can make a few general statements about such potentials, which can be written as functions of $r \equiv (AB^2)^{1/3}$.  To begin with, the divergence of the potential at $A=B=0$ is a desirable feature to avoid collapse: the FRW singularity probably disappears. Also, the quantum system keeps the invariant plane at $A=B$ and $p_A=p_B/2$. The Hamiltonian describing the dynamics in the invariant plane has then the following expression:

\begin{equation}
\label{eq4c}
H(A,p_A)=\frac{3{p_A}^2}{8A}+\frac{3A}{2}-2\Lambda A^3+ \sigma V(A)
\end{equation}
\noindent with $\sigma=0$ for the classical case. The dynamics in these planes are
simply the level sets of the above Hamiltonian function.
	
The fixed points on this invariant plane can also be easily described, regardless of the particular form of $V(r)$. The introduction of the quantum potential slightly shifts the classical fixed point, $A_0$, towards smaller values and a new one, $A_1$, appears between $0$ and $A_0$. While the former is always a saddle-center, the latter can be either a center-center,  if 
\begin{equation}
\left.\frac{d^2 V}{d A^2}\right|_{A_1} > \frac{3A_1}{\sigma A_0^2},
\label{crit}
\end{equation}
or a saddle-center otherwise. It is clear that an equality in the above equation would correspond to a bifurcation point. We recall that $A_0$ is the classical saddle-center fixed point, given by (\ref{eq4}).

\begin{figure}[h]
\begin{center}
\epsfig{file=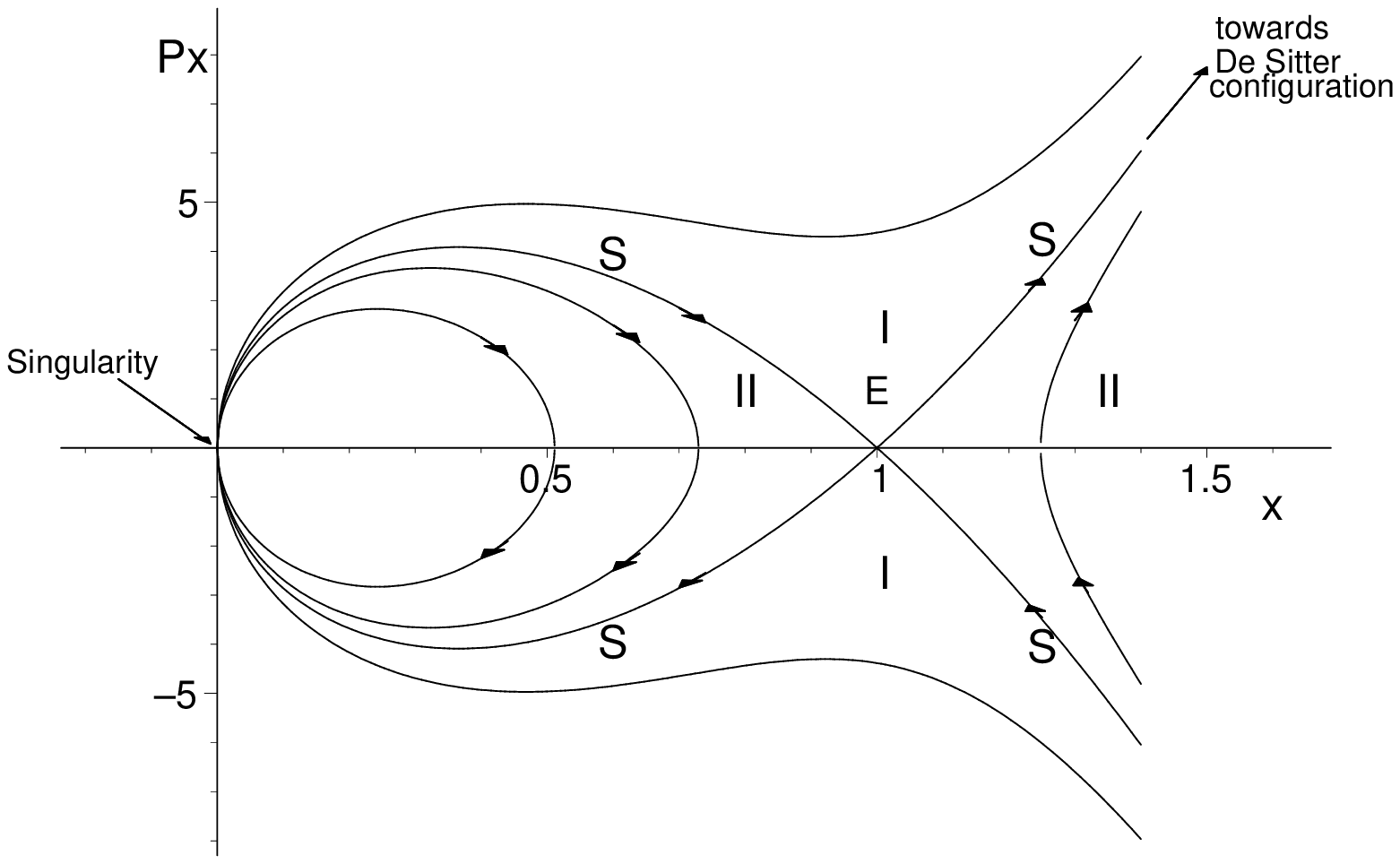,width=8cm}
\epsfig{file=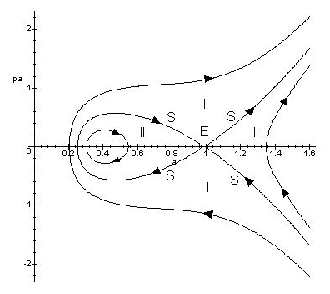,width=6cm}
\end{center}
\caption{Phase space portrait of the invariant plane $A=B$, $p_A=p_B/2$.  a) The separatrices $S$ are characterized by the energy $E_0=E_{crit}=1/\sqrt{4 \Lambda}$. The point $A=B=0$ is a degenerate singularity; b) The invariant plane for $\sigma \ne 0$: note that the quantum term removes the singularity $A=B=0$ and the trajectories bounce back in the vicinity of this point. }
\label{planinv}
\end{figure}
Moreover, for both cases, a straightforward analysis of the infinity of phase space shows that it has two \textit{attractors} in this region. One acts as an attractor (stable configuration) and the other as a repeller (unstable configuration). The scale factors, $A$ and $B$, in the classical case ($\sigma=0$) approach the attractor 
as $A=B \sim$~exp~$(\sqrt{\Lambda/3}t)$ and $p_{B}=2p_{A} \sim$~exp~$(2 \sqrt{\Lambda/3} t)$ (cf. Figure 1).

In order to further analyze the problem, we have to choose a especific form for $V(r)$ and to pick particular values for the free parameters $\sigma$ and $\Lambda$. We make a direct generalisation of (\ref{eq9a}) and write
\begin{equation}
V[r\equiv(AB^2)^{1/3}]=\frac{1}{r^2}.
\end{equation}
Taking the pair $\Lambda=1/4$ and $\sigma=0.01$, which will be used throughout this work,  we also find the saddle-center equilibrium point $E$ , slightly dislocated (as expected):
\begin{equation}
\label{eq4a}
E~: ~p_A = p_B = 0, \quad A_0= B_0=0.99317156, 
\end{equation}
with  associated energy $E=1.001006820$, higher than in the classical case. The
 eigenvalues for this equilibrium point are  ${\lambda^q}_{1,2}=\pm 0.49479869$ and
${\lambda^q}_{3,4}=\pm 1.42393682i$, so that  $E$ continues to be as saddle-center. 
However, the inclusion of the quantum potential generates another equilibrium point which is not a collapse given by
\begin{equation}
\label{eq4b}
\bar E~: ~p_A = p_B = 0, \quad A_0= B_0=0.24194223, 
\end{equation}
and the corresponding eigenvalues are  ${\lambda^q}_{1,2}=\pm 2.4044353$ and
${\lambda^q}_{3,4}=\pm 5.84525307i$ which characterize this equilibrium point as a center-center, in agreement with the criterium given at (\ref{crit}).  This is the first significant change that the quantum term brings to the dynamics of the models. The corresponding consequences will be seen in Sec.~\ref{homoclin}.

As one can see in the invariant plane shown in Fig.~(\ref{planinv}) due to the new fixed point, the left unstable branch of the separatrix connects to the left stable one, creating a homoclinic loop and avoiding the collapse.
  
In the vicinity of the saddle-center $E$  it is always possible to find a set of canonical variables such that the Hamiltonian may be expressed as \cite{moser1} 

\begin{eqnarray}
\label{eq9}
\nonumber
H(q_{1},q_{2},p_{1},p_{2})&=&\frac{\sqrt{\Lambda}}{2}(p_{2}^{2}-q_{2}^{2})
-\sqrt{2 \Lambda}(p_{1}^{2}+q_{1}^{2})\\ 
&+&{\cal{O}}(3)+\Delta E=0
\end{eqnarray}
where $\Delta E=(E_{crit}-E_{0}).$ The critical point $E$ is located at the origin $q_{1}=q_{2}=p_{1}=p_{2}=0$, with $E_{crit}=E_{0}$. Here ${\cal{O}}(3)$ denotes 
higher-order terms in the expansion.  In the approximation where ${\cal{O}}(3)$-terms are neglected, the Hamiltonian is separable with the approximate constant of motions given by the partial energies $E_{2}=\sqrt{\Lambda}(p_{2}^{2}-q_{2}^{2})/2$ and $E_{1}={\sqrt{2 \Lambda}}(p_{1}^{2}+q_{1}^{2})$. Note that $E_{1}$ is always positive. In a small neighborhood of $E$ the motion will be the composition of rotational motion and hyperbolic motion connected respectively to the conserved quantities $E_{1}$ and $E_{2}$. To display the topology of the linearized motion in a neighborhood
   of $E$ we consider the two following possibilities.

If $E_{2}=0$ and $p_{2}=q_{2}=0$ the motion corresponds to linear unstable periodic orbits $\tau_{E_{0}}$ in the plane $(q_{1},p_{1})$. Such orbits depend continuously on the parameter $E_{0}$. The second possibility is $E_{2}=0$ and $p_{2}=\pm q_{2}$, that defines the one-dimensional linear stable $V_{S}$ and linear unstable $V_{U}$ manifolds. At the critical point $E$, these manifolds are tangent to the separatrices $S$ of the invariant plane. The direct product of the periodic orbit $\tau_{E_{0}}$ with $V_{S}$ and $V_{U}$ generates, in the linear neighborhood of $E$, the structure of stable $(\tau_{E_{0}}\times V_{S})$ and unstable cylinders $(\tau_{E_{0}} \times V_{U})$. Every orbit which constitutes these cylinders coalesce to the orbit $\tau_{E_{0}}$ for times going to $+ \infty$ or $- \infty$, respectively. 
The energy of any orbit on the cylinders is the same as that of the periodic orbit $\tau_{E_{0}}$. Two types of nonlinear extension can be considered for this separable linearized motion. The first is the nonlinear extension of the cylinders, away from the nonlinear periodic orbit, which will be the subject of Section IV. The second is the nonlinear extension of the rotational motion plane, that we start to discuss
now. 

In fact, as shown in the next section, in the nonlinear regime (when higher order terms ${\cal{O}}(3)$ of the Hamiltonian are taken into account) the plane $(q_{1},p_{1})$ of the rotational motion extends to a two-dimensional manifold, the center manifold, of unstable periodic orbits of the system \cite{carr}. 
The intersection of the center manifold with the energy surface $H(E_{0})=0$ is a periodic orbit from which a pair of cylinders emanates. The structure of the center manifold of unstable periodic orbits and the associated cylinders for both Hamiltonians (\ref{eq2}) and  (\ref{eq2a}) will be displayed  in the next Sections, through a rather detailed numerical treatment. Finally we must remark that from (\ref{eq9}) the center manifold of unstable peridic orbits is defined only for $\Delta E \geq 0$.

\section{The Center Manifold of Unstable Periodic Orbits}

In order to study the full dynamics it is necessary to first calculate the periodic orbits of    the full system based on the preceeding linear analysis. These orbits can be continued to the periodic orbits of the full system by the  numerical continuation technique of computing the family of $\tau_{E_0}$ periodic orbits \cite{simo}\cite{stuchi1}. 
In general, to compute a T--periodic orbit of an autonomous system $\dot x=f(y)$, through $x_0$, means to find the zeros of the function $F(x_0)=\varphi(T,x_0)-x_0=0$, where $\varphi(t,x_0)$ is the solution starting at $x_0$ when $t=0$. It is equivalent to dealing with the problem of finding a fixed point of a Poincar\'e map ${\cal P}(x_0)$ of $\varphi(t,x_0)$ associated to a section $\Sigma$ transversal to the flow. Given $\Sigma$ and ${\cal P}(x_0)$ as above we have now to find $x_0$ $\in$ $\Sigma$ such that
\begin{equation}
\label{10}
{\cal F}(x_0)=(\varphi(t,x_0) \cap  \Sigma) -x_0={\cal P}(x_0)-x_0=0.
\end{equation}
Equation (\ref{10}) can be transformed into a numerical continuation
problem which permits to follow the periodic orbits along a desired family. A convenient Poincar\'e section is $p_B=0$, $\dot p_B <0$. Throughout this paper we fix $\Lambda=1/4$ and $\sigma=0$ and $\sigma=0.01$, such that $E_{crit}=1.0$ and $E_{crit}= 0.99317156$ for the classical quantum  cases, respectively. 
 We start to determine the $E_0$-parametric family $(0 \le E_0 \ge 1)$  of non-linear 
periodic orbits continued from the linear approximation around $E$. The predictor-corrector method uses the solution of the variational equations atthe trial solution taken from the linear approximation, reduced to the section $p_B=0$; and the differential of this Poincar\'e mapping is checked for area preservation at each step.

\begin{figure}
\begin{center}
{\includegraphics[width=6.0cm]{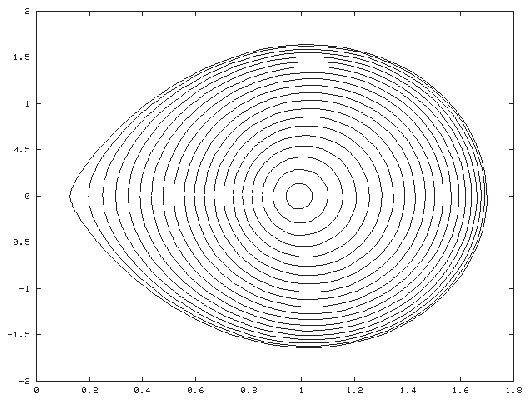}}\\ 
\includegraphics[width=6.0cm]{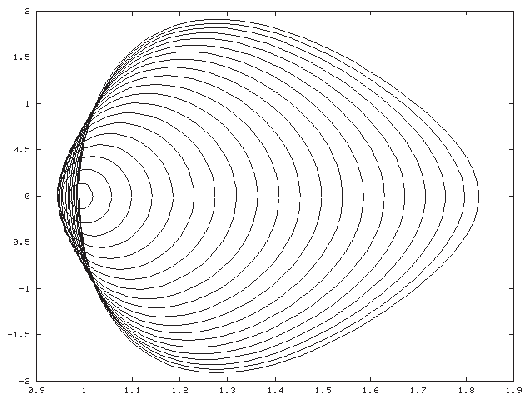}\\
\end{center}
\caption{Projection of the center manifold with $\sigma=0.01$:  (top)
 $(A,p_A)$ plane; (bottom) $(B,p_B)$ plane.  Note that the  periodic orbits
are qualitatively similar to classical ones.}
\label{cenquan}
\end{figure}

 In Figure \ref{cenquan}~(top) we show both the $(A,p_A)$ and $(B,p_B)$
 projections for the quantum case for $\sigma=0.01$. The shape of the center manifold 
is similar in both cases (classical and quantum), and as in the classical case, the periodic orbits avoid $B=0$. 
 
The above pictures show that the center manifold is a shell-like surface in phase space. From this shell  the stable and unstable  manifolds arrive/emanate asymptotically. They act as guides for nearby trajectories either taking them to the neighborhood of the center manifold or away from this neighborhood towards other regions of the phase space. An important question we want to answer is whether these manifolds collapse and cross either $A=0$ or $B=0$ before being able to  escape to the attractor like in the classical case discussed in \cite{stuchi2, uggla}. In order to see this we want to find where (in a convenient Poincar\'e section) there is a first intersection of the stable and stable manifolds  (if they indeed do cross) signalizing homoclinic points of this map which correspond to homoclinic orbits of the full dynamics. The existence of such points in the physically meaningful region  will be discussed in the next section together with a summary of the methodology. In the classical case, such homoclinic points occur only in the non-physical region, where $A<0$ \cite{stuchi2, uggla} since $B$ is never negative. 

\section{Homoclinic Intersections of the Unstable and Stable Manifolds}
\label{homoclin}

The extension of the linear unstable/stable manifolds can cross each other
 creating homoclinic orbits and a chaotic region dynamics, and indeed this is  the 
case  as our subsequent numerical analysis shows. We will see that the addition of the quantum potential brings the stable and unstable manifolds to the the physically significant region as opposed to the classical case \cite{stuchi2}.
This means that the dynamics of the homoclinic tangle can fully act  on the question for the Bianchi IX models with an ``ad hoc" quantum potential: how many times a trajectory bounces before escaping to the de Sitter region.

\begin{figure} 
\begin{center}
{\includegraphics[width=7cm]{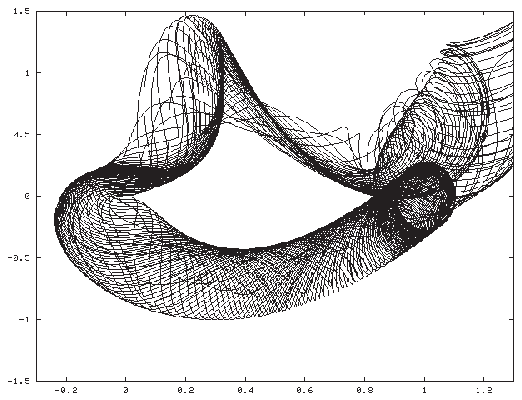}}\\
{\includegraphics[width=7cm]{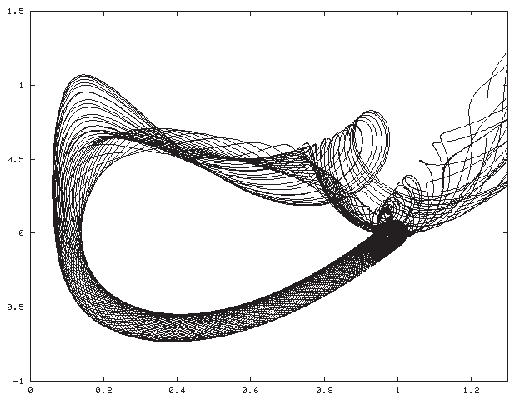}}
\end{center}
\caption{$(A,p_A)$ projections of unstable cylinder manifolds: branch of the unstable manifold  which goes towards $A=0:$ for (top) $\sigma=0$ and (bottom) $\sigma=0.01$. Note that the addition of the quantum potential makes the unstable  manifold avoid the collapse by bouncing back towards the singularity.}
\label{tube1}
\end{figure}

\begin{figure} 
\begin{center}
{\includegraphics[width=7.0cm]{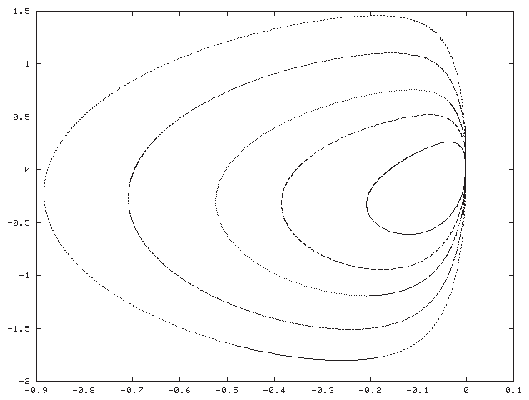}}\\
{\includegraphics[width=7.0cm]{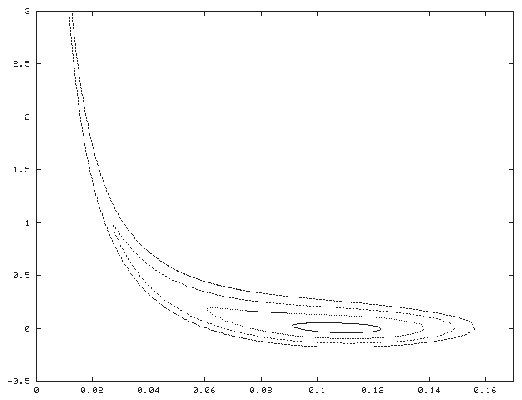}}
\end{center}
\vspace{-.5cm}
\caption{From top to bottom, (top) $\sigma=0$ (classical case) four  cuts $(A,p_A)$ of the unstable and unstable cylinder  in the surface of section $(p_B=0, {\dot{p}}_{B}<0)$, for $E_0=0.9706090$, $0.90620828$, $.70789778$ and $0.56639600$; (bottom) four cuts as above for $\sigma=0.01$ for the $E_0=1.0095581$, $1.00706088$, $1.00274454$ and $0.99685469$. Note that the quantum term brings the  Poincar\'e section of the tubes to the $A > 0$  region while they are in the $A<0$ region in the classical case (see top figure).}
\label{CORTE1}
\end{figure}

\begin{figure} 
\begin{center}
{\includegraphics[width=7.0cm]{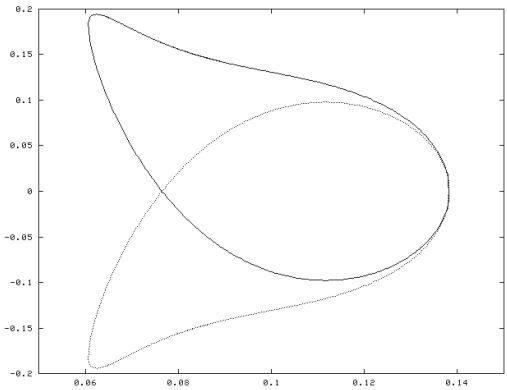}}
\end{center}
\vspace{-.5cm}
\caption{$(A,p_A)$ first iteraction of the Poincar\'e map in the surface of section 
$(p_B=0, {\dot{p}}_{B}<0)$ of  orbits on the unstable and stable cylinder manifolds
to the p.o. with $E_0=1.00706088$. Note that they are mirror images of each other with respect to $p_A=0$ due to the symmetry $(A,p_A,t)\leftarrow(A,-p_A,-t)$ in the Poincar\'e surface of section. The two intersections represent the 1-turn homoclinic orbits.}
\label{cut1}
\end{figure}

\begin{figure} 
\begin{center}
{\includegraphics[width=7.0cm]{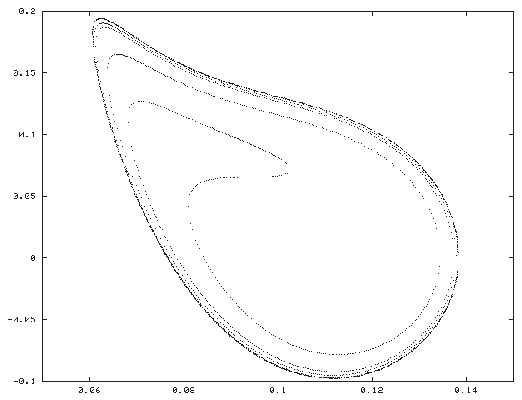}}\\
{\includegraphics[width=7.0cm]{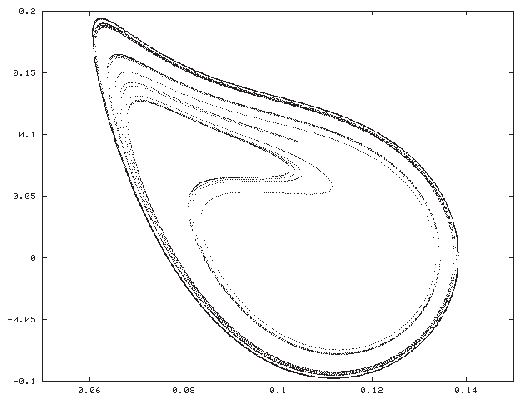}}\\
\end{center}
\vspace{-.5cm}
\caption{$(A,p_A)$ (top) second and (bottom) third iteractions of the Poincar\'e map in the surface of section $(p_B=0, {\dot{p}}_{B}<0)$ of  orbits on the unstable manifolds to the p.o. with $E_0=1.00706088$. Note that the external shape of the second cut is practically the same as the first cut shown in Fig.\ref{cut1}. The external part of the third cut mimics the second one.}
\label{cut23}
\end{figure}

\begin{figure} 
\begin{center}
{\includegraphics[width=8.0cm]{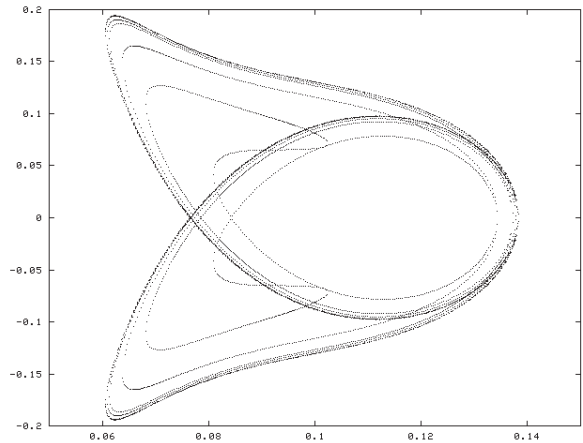}}\\
\end{center}
\vspace{-.5cm}
\caption{$(A,p_A)$ superposition of the first and second  iteractions of the Poincar\'e map in the surface of section $(p_B=0, {\dot{p}}_{B}<0)$ of  orbits on the unstable and stable manifolds to the p.o. with $E_0=1.00706088$.}
\label{cut2is}
\end{figure}

\begin{figure}
\begin{center}
{\includegraphics[width=7.cm]{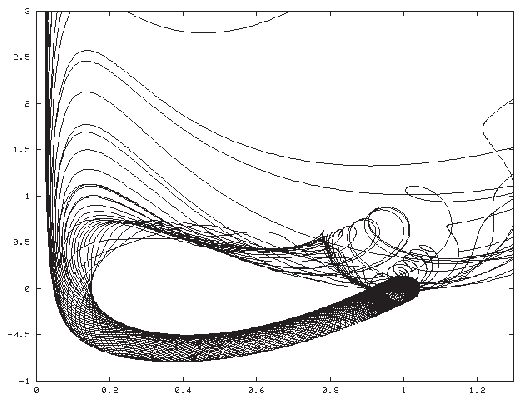}}
{\includegraphics[width=7.cm]{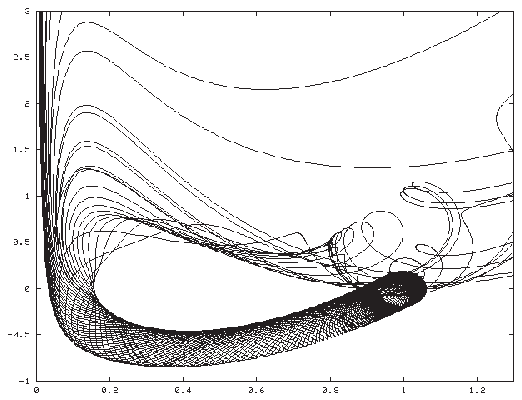}}
\end{center}
\vspace{-.5cm}
\caption{The unstable manifold branch (top) at the onset of very large values of $p_A$ at $E_0=1.00274454$; (bottom) for $E_0=0.99585469$, that is, for a larger value of energy  the density of  trajectories that approach $A=0$  increases, but part of them approach the periodic orbit and escape to the de Sitter region.}
\label{fuga}
\end{figure}

We first  (numerically) calculate the stable and unstable cylinders 
by a technique denoted Numerical  Globalization of Manifolds \cite{simo,stuchi1}. Second, in a Poincar\'e section transverse to a periodic orbit (here, $p_B=0, {\dot{p}}_{B}<0$ ), so that the orbit is seen as a  point and the cylinders are curves tangent to the directions of the eigenvectors of the Monodromy Matrix corresponding to the real eigenvalues.  We note that these curves divide the energy space in  interior and exterior regions and are formed by the iteractions of the orbits belonging to  an invariant manifold. Therefore,  orbits in each one of these two regions are subject to different outcomes under further iteractions if there are  homoclinic intersections,
as discussed later on.

The symmetry of the problem in the section ($(A,p_{A},t) \rightarrow (A,-p_{A},-t)$) allows saving computational effort: from the Poincar\'e map of the unstable manifold
of a periodic orbit of the center manifold, we  can obtain the corresponding Poincar\'e map of  the stable one by just applying this symmetry. The branches of the stable and unstable manifolds which go towards the origin are the ones that intersect each other. The opposite branches go to the attractor-repeller \textit{points}.

In Figure \ref{tube1}~(bottom) we show the branch of the unstable manifold, in the quantum case ($\sigma=0.01$), for $E_0=0.97112431$ which goes towards the region $A=B=0$  without collapsing, and returns again to a neighborhood of the periodic orbit from which it emanated. Some orbits escape to the attractor-repeller \textit{points} while the remaining ones return again towards $A=B=0$. This is an indicative of homoclinic chaos. In Figure \ref{tube1}~(top) we show the corresponding  branch of the unstable manifold for the classical case $(\sigma=0)$. It emerges from a periodic  orbit with energy $E_0=0.99470278$ and there are also escaping and returning trajectories. However, in the classical case this branch of the unstable manifold crosses the singularity $A=0$, indicating that the physical collapse takes place before it has time to return to the neighborhood of the periodic orbit. The same is true for the stabe manifold, which in this projection is mirror symmetric with respect to the $A-axis$ (not shown here to avoid cluttering the figures). So far the non collapsing branches of these manifolds are the main difference between classical and quantum models. Note that in the latter case, the singularity is avoided. This feature could have been guessed from the difference between the invariant planes shown in Figures~\ref{planinv}~(top) and (bottom).

In Figure \ref{CORTE1}~(top) and (bottom) we show four cuts of  unstable manifolds in the surface of seccion ($p_B=0,\dot{p_B}< 0$) for the classical case ($\sigma=0$) and quantum case ($\sigma=0.1$), respectively, for decreasing values of the energy as shown in the label of the Figure. The introduction of the quantum potential changes radically the scenario: the cuts are brought to  the physical significant region $A >0$. We note that as the trajectories of the tubes going towards $A=0$ the momentum $p_A \rightarrow \infty$. However for energies near the critical point  the stretching effect is not so strong.  In the classical case it was shown in \cite{stuchi2} that there are two homoclinic orbits that are seen in the Poincar\'e section as the intersection of the two curves representing the unstable manifold which occurs at $A<0$. As can be seen in Figure \ref{cut1} where we show both the unstable and stable manifolds sections, there are two points of intersection: one near $A=0.14$ and the other near $A=0.8$. The energy of the periodic orbit, as well as of the stable/unstable manifolds is $E=1.00706088$. Note that the sections of these two manifolds are mirrors of each other as commented above. The two intersections represent homoclinic orbits comming from the periodic orbit through the unstable manifold, making one turn and returning assymtoptically to the periodic orbit through the stable manifold.  Figures \ref{cut23}~(top) and (bottom) exhibt the  second  and third cuts of the unstable manifold, respectively. As can be seen in the second cut the set of orbits which do not escape form a very convoluted figure which gets larger and thinner as it tends to the first cut from its inside. The third cut which
can be seen in Figure \ref{cut23}~(bottom) is even more convoluted since it has to goes towards the second  cut: note that in this figure the "contour" of the first cut is denser due to the accumulation of points from the second and third cut. We also observe that the third cut is folded back inside the second one. Figure \ref{cut2is}  shows two cuts of the unstable and stable manifolds and it is clear that higher-order homoclinic points have been created besides the first two ones shown in Figure~\ref{cut1}. These orbits are asymptotic to the periodic one after two turns. In this way a whole mesh of homoclinic points and initial conditions on different sides of these points can lead to escape after a given number of bounces. 

Therefore it is clear that the addition of the quantum potential creates a homoclinic tangle  contained  in the $A$-and-$B$-greater-than-zero region, meaning that the escaping ruled by this homoclinic tangle has no chance to collapse.  However, as the energy decreases the cuts go near $A=0$ and become more elongated (see Figure \ref{fuga})  and the $p_A$ moment becomes very large.  This feature needs to be further examined. Note that the number of orbits having very large moment before turning towards the periodic orbit increases with decreasing total energy. 
 
\section{Final comments and perspectives}

We study one case of a class of quantum-inspired potentials in the classical axisymmetrical Bianchi IX models, namely $V[r\equiv (AB^2)^{1/3}] $. After pointing out the general features of such potentials, we  focus on a particular expression, similar to the one exactly calculated for FRW model \cite{lemos} which  is equivalent to adding a repulsive potential near the origin. The effect of this quantum term is to avoid the singularity and to give rise to an inflationary escape to the de Sitter region  ruled by a chaotic scattering in opposition to the classical case where the two possible outcomes are separated only by the stable/unstable manifolds without the richness of the homoclinic tangle of these manifolds.

\acknowledgments
 We  thank Dr. I. Dami\~ao Soares for suggesting this problem. We are very much 
indebted to Dr. N.A. Lemos for invaluable discussions. T.J. Stuchi thanks CNPq/Brazil for partial financial support. G. Corr\^ea acknowledges support from CAPES/Brazil.

\end{document}